\documentclass{PHYEAUTH}

\usepackage[dvips]{graphicx}
\usepackage[]{amsmath}
\usepackage[]{amssymb}
\usepackage[]{amsfonts}
\usepackage[]{latexsym}
\usepackage[]{float}

\begin{document}

\begin{frontmatter}

\title{
Topological low-energy modes in $N=0$ Landau levels of graphene:\\
      a possibility of a quantum-liquid ground state
}

\author[address1,address1a]
{Yasuhiro Hatsugai\thanksref{thank1}},
\author[address2]{Takahiro Fukui} and
\author[address3]{Hideo Aoki}

\address[address1]{
Institute of Physics, University of Tsukuba, 
Tennodai, Tukuba 305-8571
Japan
}
\address[address1a]{
Department of Applied Physics, University of Tokyo, Hongo, 
Tokyo 113-8656, Japan
}

\address[address2]{
Department. of Mathematical Sciences, Ibaraki University, Mito 
310-8512, Japan
}

\address[address3]{
Department of Physics, University of Tokyo, Hongo, 
Tokyo 113-0033, Japan
}

\thanks[thank1]{
Email:hatsugai@sakura.cc.tsukuba.ac.jp
}

\begin{abstract}
We point out that the zero-energy Landau level of Dirac fermions 
in graphene can be, in the presence of a repulsive electron-electron
interaction, split into two associated with a ``bond ordering" formation 
having a ``Kekul\'{e} pattern'', which respects the chiral 
symmetry.  Since the Kekul\'{e} pattern has a three-fold degeneracy, 
domain structures are implied, for which we show that 
in-gap states localized along the domain boundaries exist 
as topological states.  Based on this a possibility 
of a quantum-liquid ground state 
of graphene in magnetic fields is discussed.

\end{abstract}

\end{frontmatter}

\section{Introduction}
Graphene, for which an unconventional quantum Hall effect
has recently been discovered,\cite{graphene} 
is a peculiar condensed-matter
realization of massless Dirac fermions.
While the quantum Hall effect is generally characterized by
a topological (Chern) number, the peculiarity of 
Landau levels (LL) in graphene appears as 
 existence of the exactly zero energy  Landau level. 
We have established topological aspects in the graphene QHE\cite{HFA06}, 
among which is 
a bulk-edge correspondence coming from the topological 
nature of bulk and edge states.
All the topological features are intimately related with 
the chiral (bipartite) symmetry of the honeycomb lattice,
which, in addition to being responsible for the zero energy 
 Landau level, 
gives a characteristic
coexistence of extended and edge states at $E=0$\cite{HFA06}.

Now, an interesting question is: can the zero energy 
Landau level of the Dirac fermion's
be lifted when 
$E_F$ is situated right in the zero-th Landau level.  
In the present paper we propose that an intriguing candidate 
for triggering the 
splitting should be a ``bond ordering'', 
in which the electronic bonds have an ordering 
with a 
``Kekul\'{e} pattern'', arising either from a Jahn-Teller type distortion or
electronic bond-order 
($\langle c^\dagger_ic_j\rangle $) formation due to 
electron-electron interactions\cite{HFA07}.  
Reflecting the hexagonal symmetry there are three equivalent 
ordering (``Kekul\'{e}'') patterns (see Fig.1), and the bond 
ordering is one way to break the degeneracy. 
The ordering, when static, produces a mass in 
the Dirac fermion's dispersion (in zero magnetic field), 
despite the fact that 
the bond ordering does {\it not} break the chiral symmetry.
In magnetic fields, the bond ordering acts to 
split the zero-energy Landau level of the 
Dirac fermions, which may be viewed as a kind of the Peierls instability 
since the one-particle density of states diverges. 

\section{Bond ordering  and the split zero-energy Landau levels}
We now examine the above idea in the simplest manner, 
i.e., we assume that the ordering is static and 
adopt a Hamiltonian that has the hopping $t_{ij}$ with 
a Kekul\'{e}-type modulation, which represents, in a mean-field sense, 
the bond ordering, say, 
 originating from the electron-electron interaction.
We first take the case in which 
the ordering pattern has a translational symmetry as in Fig.1. 
The 
massless Dirac fermions then acquire a mass which opens a gap 
(Fig. 2).  As stressed the bond ordering does not destroy the 
chiral symmetry (although the size of the Brillouin zone is expanded), 
which implies that the particle-hole symmetry is preserved as well for any 
(including random) ordering patterns.  
This contracts with other types of modulation (such
as site dependent potentials) which break the 
chiral symmetry.  Correspondingly, the zero-energy 
Landau level in uniform magnetic fields, which is a 
special Landau level sitting right at the Dirac cone vertex,
also splits into two  
\cite{nakajima}. 

This implies that the bond ordering, 
which is here assumed, is expected to spontaneously induced 
in a self-consistent treatment as a kind of Peierls instability 
in the zero-energy LL\cite{HFA07}.

\begin{figure}[h]
\begin{center}
 \includegraphics[width=0.6\linewidth]{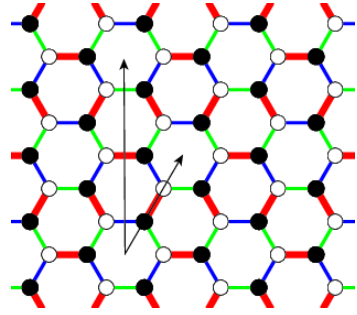} 
\end{center}
\caption{Kekul\'{e} (static bond-ordering) patterns 
have three, equivalent realizations, where either the red, 
blue or green bonds are $r(>1)$ times stronger than the others. 
The two arrows are unit vectors of the enlarged unit cell. (color online)
}
\label{f:disp}
\end{figure}

\begin{figure}[htb]
\begin{center}
\includegraphics[width=0.48\linewidth]{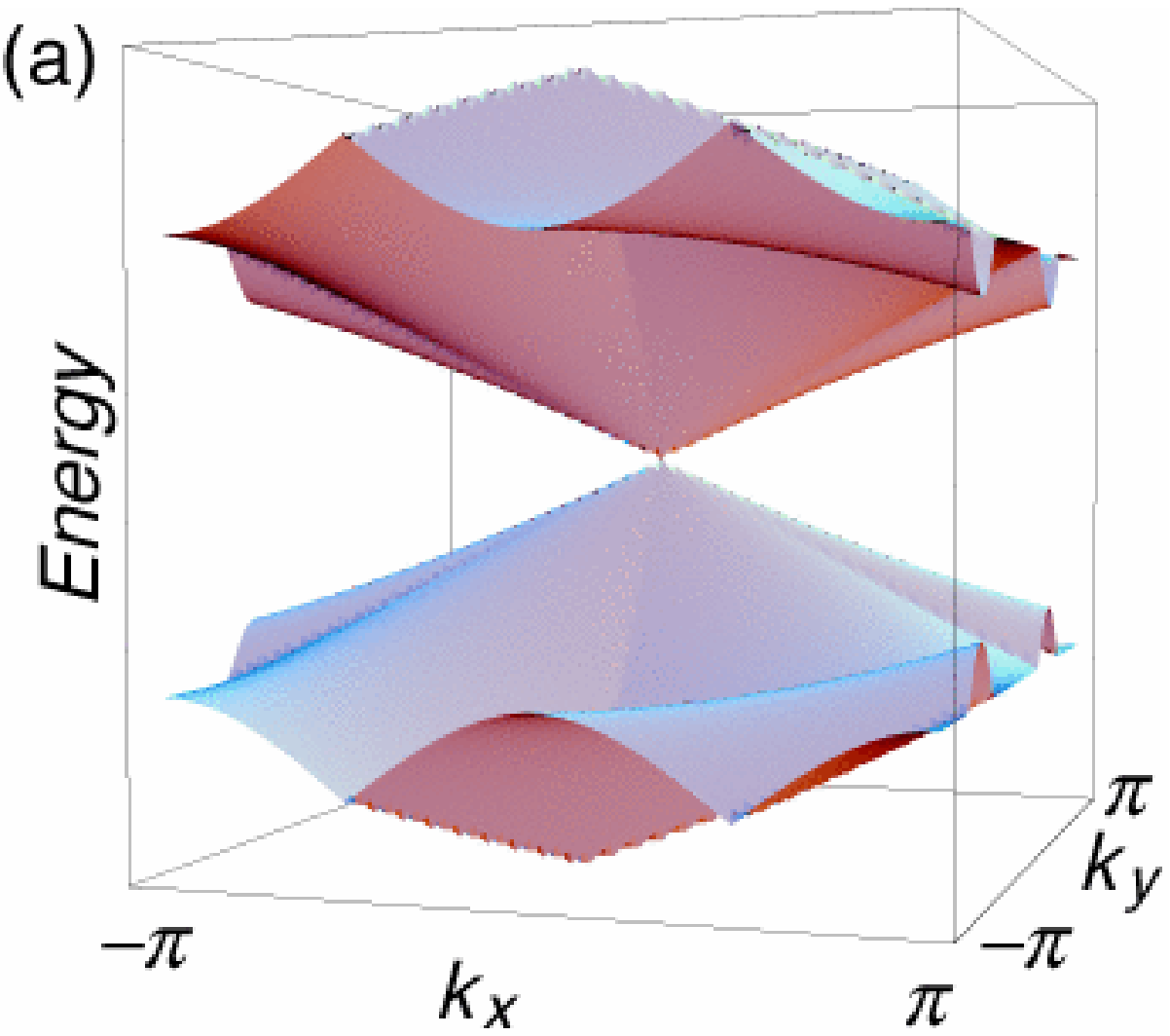}
\includegraphics[width=0.48\linewidth]{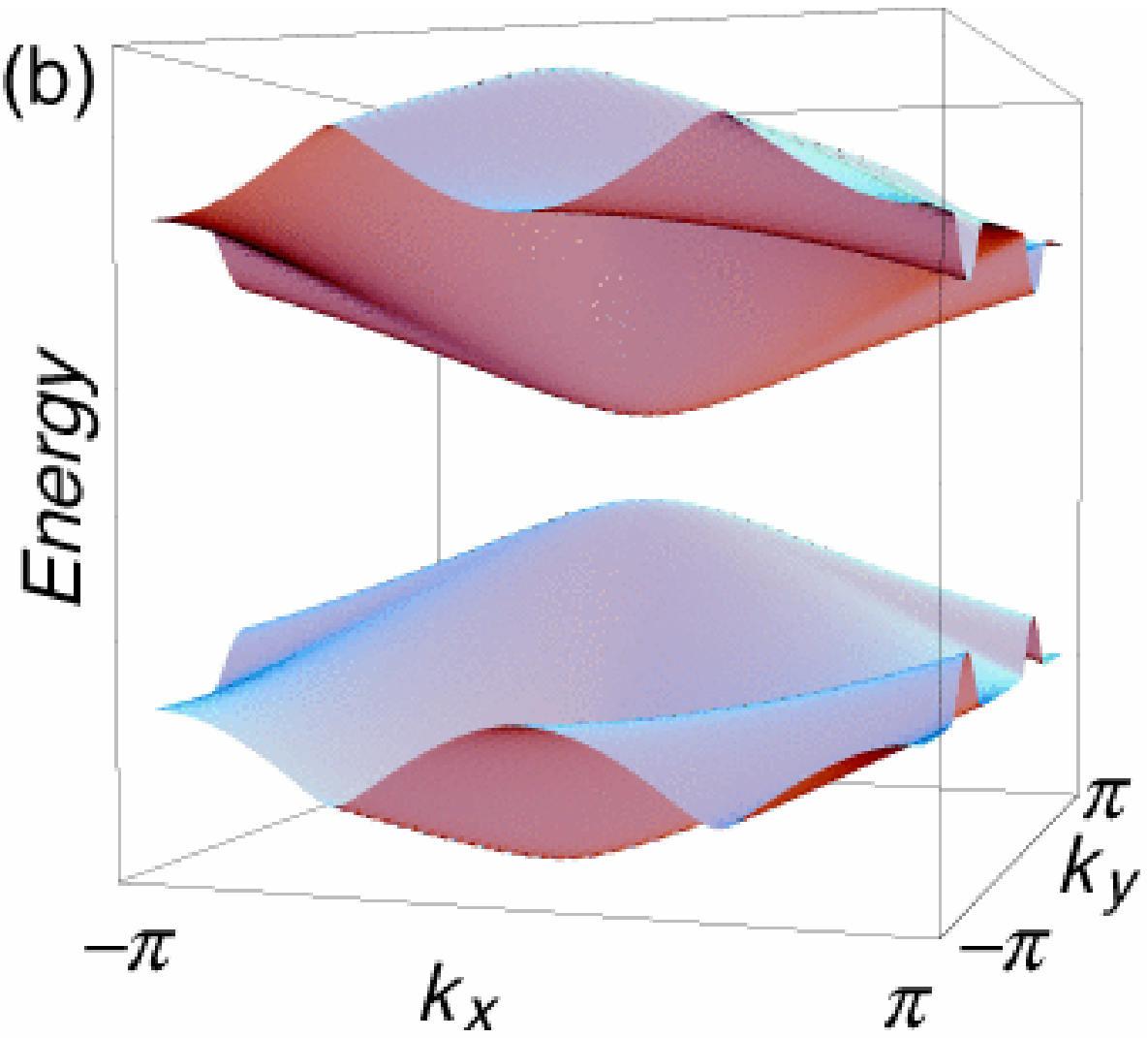}
\end{center}
\caption{(a) 
The band dispersion around $E=0$ 
in graphene before (a) and after (b) 
a (static) bond ordering is introduced, 
as displayed in the folded Brillouin zone 
for $r=1.2$ here.}
\label{f:disp}
\end{figure}

\section{Topological states localized along domain boundaries 
as 2D analogue of solitons}
We have seen that the bond ordering, when translationally symmetric, 
splits the zero-energy Landau level accompanied by 
an energy gap at $E=0$.  
However, there are a vast number of possible 
realizations of bond-ordering patterns that have no 
translational symmetry.  
Specifically, there is an important class of such states that consist 
of ``domains".  Namely, there are three, equivalent 
`Kekul\'{e} patterns (differently colored in Fig. 1), 
and there are a huge number of configurations where, 
e.g., a bond-ordered phase with strong red bonds 
sits next to another phase with strong blue bonds (Fig. 3).  
We can then expect that ``boundary states" 
that are localized along the {\it domain boundaries} should appear
as generalized boundary states in graphene, 
which have topological origin and stability.\cite{HFA06}  
The situation reminds us of the well-known {\em soliton} modes 
across the different conjugated patterns in 1D polyacetylene.  
So the boundary states considered here is a 
2D extension of solitons
which are
topologically protected.

The boundary states, whose energies  
reside in the gap of the split $N=0$ Landau levels, 
still do not violate the chiral symmetry.
Physically the situation is similar to the appearance of 
the zero-mode edge states along the zig-zag edges in 
graphene in zero magnetic field, 
which also have a topological origin.  
These loop-like boundary states (``strings'') may have different 
realization in 2D as further exemplified in Fig. 4. 
In each case topologically stable boundary states arise near the 
zero energy as displayed in Figs. 3(b),4(b).

\begin{figure}[h]
\begin{center}\leavevmode
\includegraphics[width=0.85\linewidth]{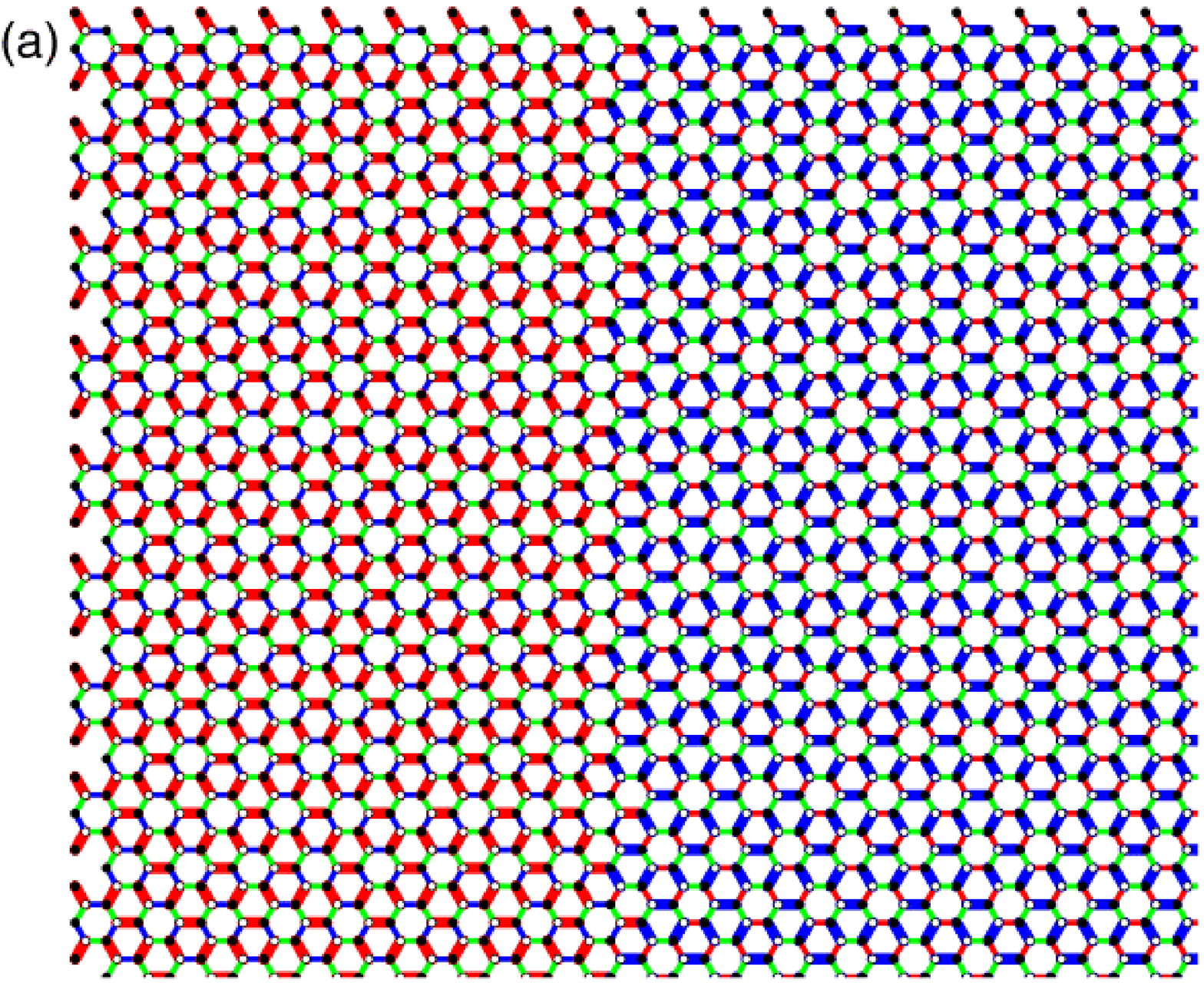}
\vskip 0.1cm
\includegraphics[width=0.85\linewidth]{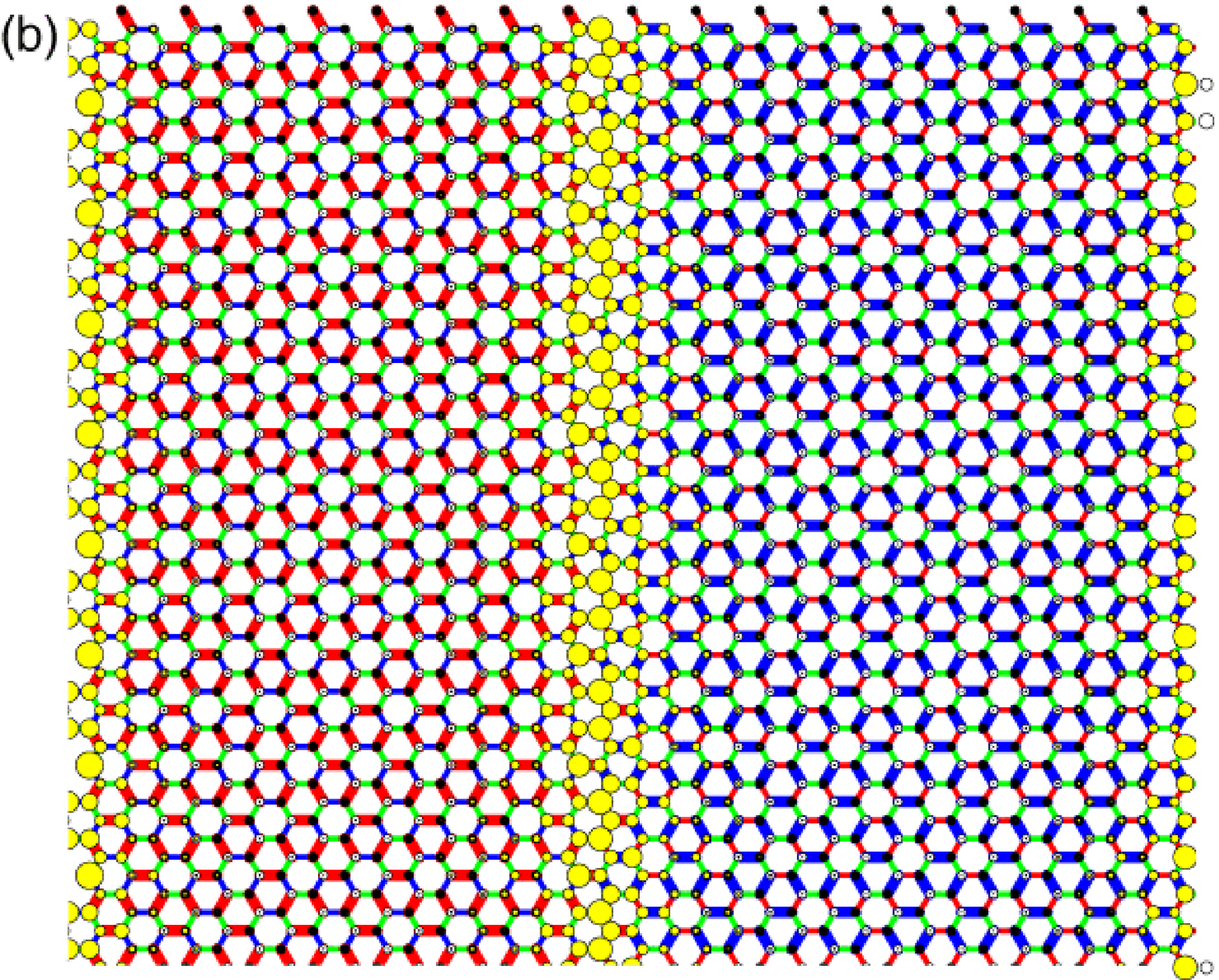}
\caption{
(a)
A bond ordering with straight domain boundaries, 
where the red (blue) hopping is stronger than the others (by a factor of 1.2) 
in the left (right) domain.  
(b) The boundary state whose energy sits between 
the split $N=0$ LL is shown with the charge density 
represented by yellow circles 
for the magnetic flux of $1/6$ times the flux quantum per 
hexagon. (color online)
} \label{fig3}
\end{center}
\end{figure}

In reality the bond ordering configuration can be 
dynamical, which may be described in terms of the 
fluctuating strings\cite{HFA07}.  
We can then conjecture that 
the true state may possibly be a quantum liquid realized 
as a condensate of such ``strings". 
The physics may have an analogy 
with the ``string net condensation" considered for spin models\cite{string}. 
A self-consistent treatment of the effect of the 
electron-electron interaction will be given elsewhere\cite{HFA07}.

This work has been supported in part by 
 Grants-in-Aid for Scientific Research on Priority Areas from MEXT, 
``Physics of new quantum phases in superclean materials"
 (Grant No.18043007) for YH, 
``Anomalous quantum materials" (No.16076203) for HA, 
and a Grant-in-Aid for Scientific Research  
(No.18540365) from JSPS for TF.

\begin{figure}[h]
\begin{center}\leavevmode
\includegraphics[width=0.85\linewidth]{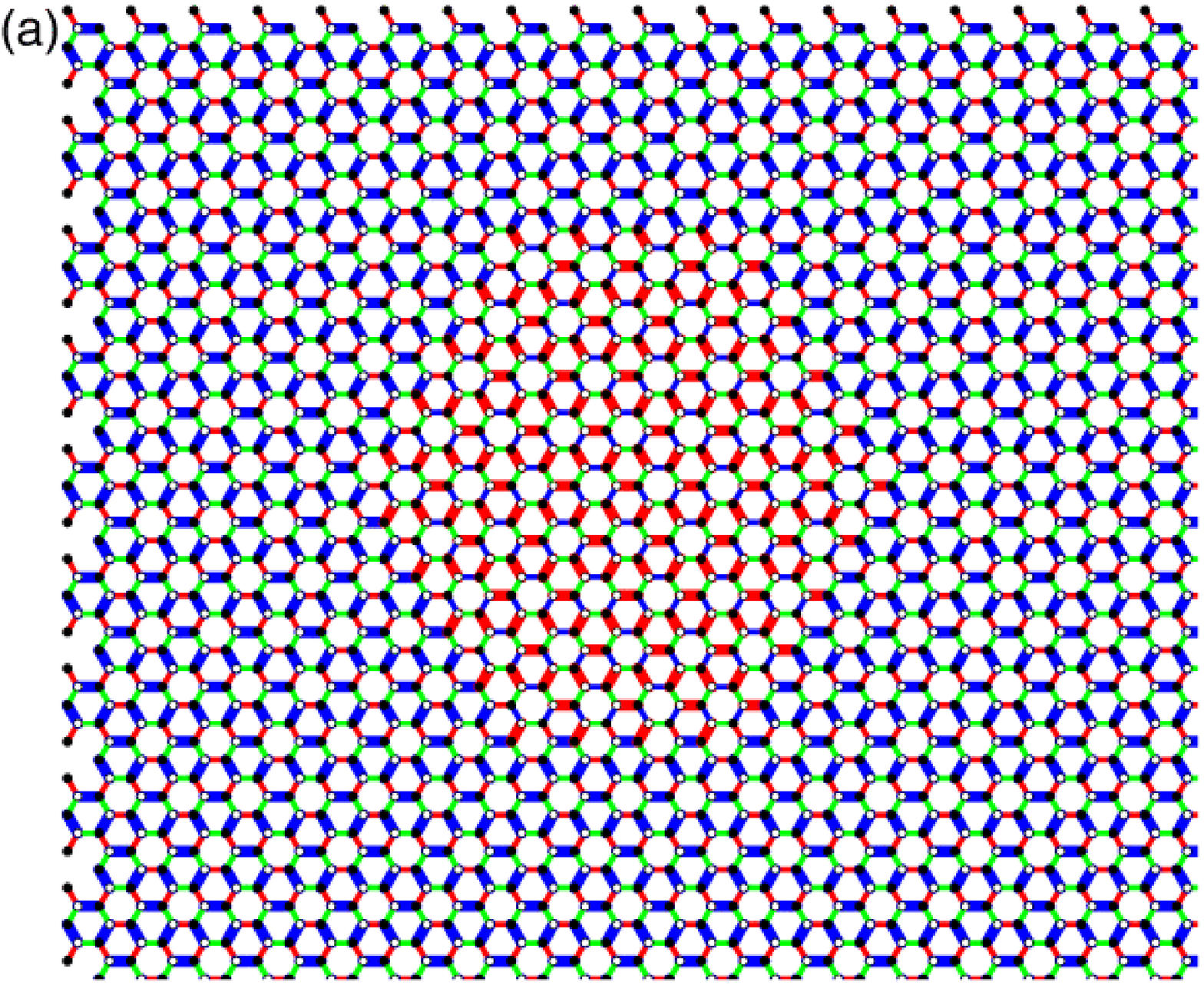}\ 
\vskip 0.1cm
\includegraphics[width=0.85\linewidth]{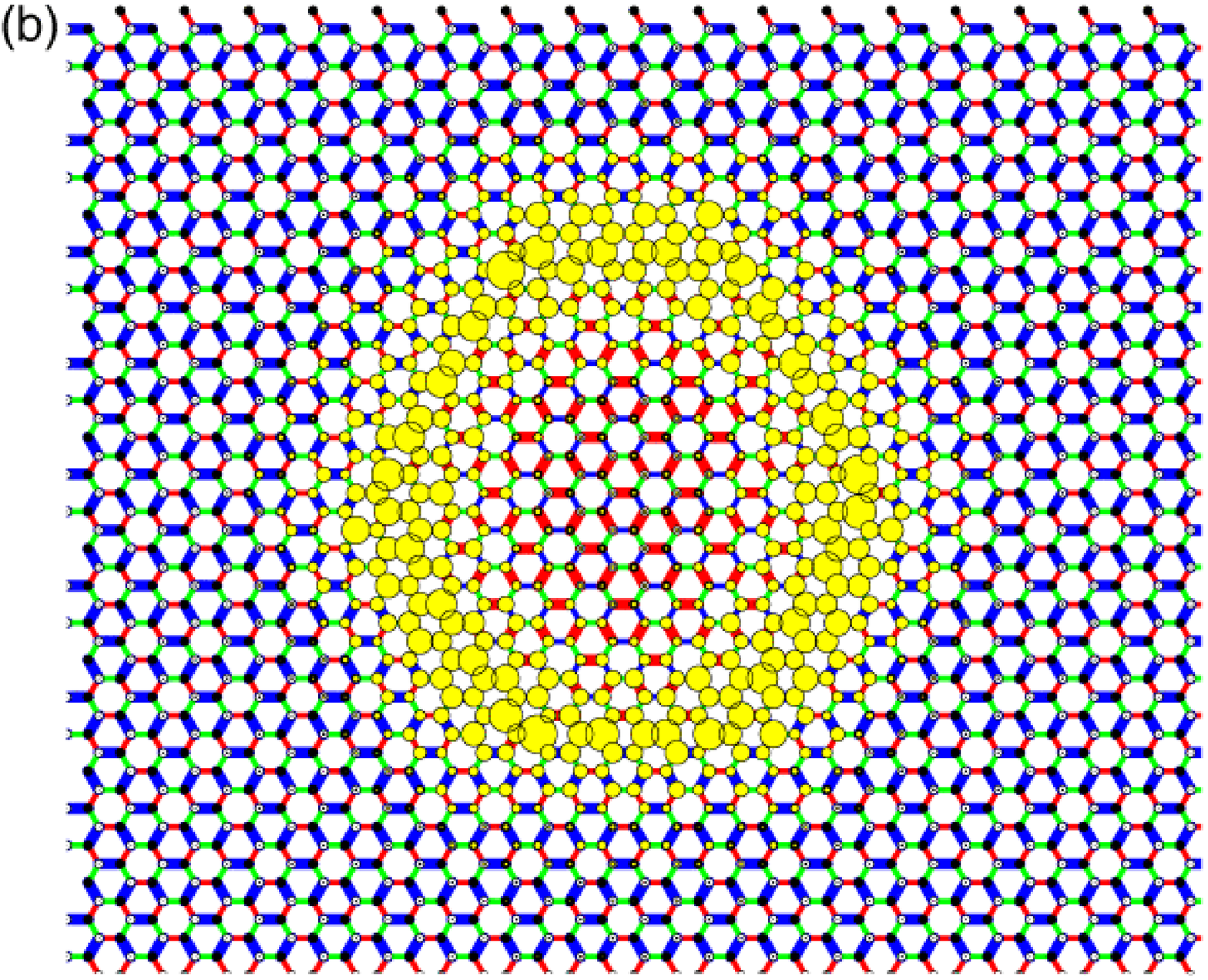}
\caption{
A bond-ordering pattern with a closed-loop boundary, 
where the red (blue) hopping is stronger than the others (by a factor of 1.2) 
in the inner (outer) domain.  
(b) The boundary state shown as the charge density 
represented by yellow circles 
for the same magnetic flux as in Fig.3. (color online)
} \label{fig4}
\end{center}
\end{figure}


\end{document}